\documentclass[aps,twocolumn]{revtex4}
\usepackage{graphicx}
\usepackage{epsfig}
\usepackage{dcolumn}
\usepackage{amsmath}
\def\Journal#1#2#3#4{{#1} {\bf#2}, #3 (#4)}

\def\PLB{{\rm Phys. Lett.}  B}
\def\PRL{\rm Phys. Rev. Lett.}
\def\PRD{{\rm Phys. Rev.} D}
\def\PRC{{\rm Phys. Rev.} C}

\def\JPG{{\rm J. Phys.} G: Nucl. Part. Phys.}
\def\EPJA{{\rm Eur. Phys. J.} A}

\def\ep{\epsilon}
\def\lam{\lambda}

\def\la{\langle}
\def\ra{\rangle}

\def\al{\alpha}

\def\be{\begin{equation}}
\def\ee{\end{equation}}
\def\bea{\begin{eqnarray}}
\def\eea{\end{eqnarray}}
\begin{document}
\title{Space- and time-like form factors for $\omega\to\pi\gamma^*$ and $K^*\to K\gamma^*$
in light-front quark model}
\author{ Ho-Meoyng Choi\\
Department of Physics, Teachers College, Kyungpook National University,
     Daegu, Korea 702-701}
\begin{abstract}
We investigate space- and time-like form factors for $\omega\to\pi\gamma^*$ and $K^*\to K\gamma^*$
decays using the light-front quark model constrained by the variational principle for the QCD-motivated
effective Hamiltonian. The momentum dependent space-like form factors are obtained in the $q^+=0$ frame and
then analytically continued to the time-like region. Our prediction for the time-like form factor
$F_{\omega\pi}(q^2)$ is in good agreement with the experimental data. We also find that the space-like form factor $F_{K^{*\pm}K^{\pm}}(Q^2)$ for charged kaons encounters a zero because of the negative interference between the two currents to the quark and the antiquark.
\end{abstract}


\maketitle
The one-photon radiative decays from the low-lying vector(V) to pseudoscalar(P) mesons,
i.e. magnetic dipole $V(1^3S_1)\to P(1^1S_0)\gamma$ transitions, have been the subject
of continuous interest both theoretically and experimentally. These processes provided
a valuable testing ground to understand the internal structure of hadrons and thus to pin
down the best phenomenological model of hadrons. In our previous light-front quark
model(LFQM) analysis~\cite{CJ99,CJ07} based on the QCD-motivated effective Hamiltonian,
we have calculated various radiative $V\to P\gamma$ and $P\to V\gamma$ decay widths of
light-flavored mesons($\pi,\rho,\omega,K,K^*,\phi,\eta,\eta'$)~\cite{CJ99} and
heavy-flavored ones such as ($D,D^*,D_s, D^*_s,\eta_c, J/\psi$) and
($B,B^*, B_s,B^*_s,\eta_b,\Upsilon$)~\cite{CJ07} and found a good agreement with the experimental data.
Especially for the recent analysis of the heavy meson sector~\cite{CJ07}, we have  calculated not only
the decay widths but also the momentum dependent transition form factors in both space- and time-like regions.
However, our previous works on the magnetic dipole transitions of light-flavored mesons~\cite{CJ99}
presented only the decay widths without showing the momentum dependent behaviors of the form factors.
Particularly interesting radiative decays of light-flavored mesons may be $\omega\to\pi\gamma^*$ and
$K^*\to K\gamma^*$ processes since the form factor for $\omega\to\pi\gamma^*$ has already been measured
in the time-like region  via the decay of $\omega\to\pi^0\mu^+\mu^-$~\cite{Dz} and $K^*\to K\gamma^*$ decays
may deserve special attention in terms of $SU(3)$ flavor symmetry breaking.

The purpose of this Brief Report is to calculate the space- and time-like transition
form factors for selected $\omega\to\pi\gamma^*$ and $K^*\to K\gamma^*$ processes using our LFQM~\cite{CJ99,CJ07}
and compare with other theoretical model predictions~\cite{Ito,Card,Munz,MT,KK,YXM} as well as the available
data~\cite{Dz}. To obtain the time-like form factor $F_{VP}(q^2)$ for $V\to P\gamma^*$, we have performed the
analytic continuation from the space-like($q^2<0$) region to the physical time-like region $[0\leq q^2\leq (M_V-M_P)^2]$. We find that the time-like form factor $F_{\omega\pi}(q^2)$ obtained by analytic continuation is
in good agreement with the data. We also find that the charged $K^{*\pm}\to K^{\pm}\gamma^*$ transition form factor encounters a zero because of the negative interference between the two currents to the quark and the antiquark.

In our LFQM~\cite{CJ99,CJ07}, the momentum space light-front wave function of the ground state
pseudoscalar and vector mesons is given by
\be\label{w.f}
\Psi^{JJ_z}_{M}(x_i,{\bf k}_{i\perp},\lam_i)
=\phi_R(x_i,{\bf k}_{i\perp})
{\cal R}^{JJ_z}_{\lam_1\lam_2}(x_i,{\bf k}_{i\perp}),
\ee
where $\phi_R(x_i,{\bf k}_{i\perp})$ is the radial wave function and ${\cal R}^{JJ_z}_{\lam_1\lam_2}$
is the spin-orbit wave function obtained by the interaction independent Melosh transformation~\cite{Mel}
from the ordinary equal-time static spin-orbit wave function assigned
by the quantum numbers $J^{PC}$. The meson wave function in Eq.~(\ref{w.f}) is represented by the
Lorentz-invariant variables, $x_i=p^+_i/P^+$,
${\bf k}_{i\perp}={\bf p}_{i\perp}-x_i{\bf P}_\perp$ and $\lam_i$, where
$P$, $p_i$ and $\lam_i$ are the meson momentum, the momenta, and the helicities of the
constituent quarks, respectively. The covariant forms of the spin-orbit wave functions
for pseudoscalar and vector mesons are given in Refs.~\cite{CJ99,CJ07}.

For the radial wave function $\phi_R$, we use the same Gaussian wave function
for both pseudoscalar and vector mesons
\be\label{rad}
\phi(x_i,{\bf k}_{i\perp})=\frac{4\pi^{3/4}}{\beta^{3/2}}
\sqrt{\frac{\partial k_z}{\partial x}}
{\rm exp}(-{\vec k}^2/2\beta^2),
\ee
where ${\vec k}^2={\bf k}^2_\perp + k^2_z$ and the Gaussian parameter $\beta$ is related with
the size of meson. Here, the longitudinal component $k_z$ of the three momentum is given by
$k_z=(x_1-\frac{1}{2})M_0 + (m^2_2-m^2_1)/2M_0$ with the invariant mass $M_0$ defined by
$M^2_0=({\bf k}^2_\perp+m^2_1)/x_1 + ({\bf k}^2_\perp+m^2_2)/x_2$.
The Jacobian $\partial k_z/ \partial x$ of the variable transformation
$\{x,{\bf k}_\perp\}\to {\vec k}=({\bf k}_\perp, k_z)$
is included in the radial wave function so that the wave function satisfies the following
normalization
\be\label{norm}
\int^1_0dx\int\frac{d^2{\bf k}_\perp}{16\pi^3}
|\phi_R(x,{\bf k}_{i\perp})|^2=1.
\ee
The key idea in our LFQM~\cite{CJ99,CJ07,CJ_PLB} for mesons is to treat $\phi_R(x,{\bf k}_\perp)$
as a trial function for the variational principle to the
QCD-motivated effective Hamiltonian saturating the Fock state expansion
by the constituent quark and antiquark. The QCD-motivated effective Hamiltonian for
a description of the ground state meson mass spectra is given by
\be\label{Ham}
H_{q\bar{q}}=H_0 + V_{q\bar{q}} =
\sqrt{m^2_q+{\vec k}^2}+\sqrt{m^2_{\bar{q}}+{\vec k}^2}+V_{q\bar{q}}.
\ee
In our LFQM~\cite{CJ99,CJ07,CJ_PLB}, we use the two interaction potentials $V_{q\bar{q}}$
for the pseudoscalar and vector mesons: (1) Coulomb
plus harmonic oscillator(HO), and (2) Coulomb plus linear confining potentials.
In addition, the hyperfine interaction, which is essential to distinguish
vector from pseudoscalar mesons, is included for both cases, viz.,
\be\label{pot}
V_{q\bar{q}}=V_0 + V_{\rm hyp}
= a + {\cal V}_{\rm conf}-\frac{4\al_s}{3r}
+\frac{2}{3}\frac{{\bf S}_q\cdot{\bf S}_{\bar{q}}}{m_qm_{\bar{q}}}
\nabla^2V_{\rm coul},
\ee
where ${\cal V}_{\rm conf}=b_lr (b_hr^2)$ for the linear (HO) potential and
$\la{\bf S}_q\cdot{\bf S}_{\bar{q}}\ra=1/4 (-3/4)$ for the
vector (pseudoscalar) meson. Our variational principle to the QCD-motivated effective Hamiltonian
first evaluates the expectation value of the central Hamiltonian $H_0+V_0$ with a
trial function $\phi_R(x_i,{\bf k}_{i\perp})$ that depends on the
variational parameters $\beta$ and then varies $\beta$ until
$\la\phi_R|(H_0+V_0)|\phi_R\ra$ becomes a minimum.  Once these model
parameters are fixed, the mass eigenvalue of each meson is obtained
by $M_{q\bar{q}}=\la\phi_R|(H_0+V_{q\bar{q}})|\phi_R\ra$. More detailed procedure of
determining the model parameters can be found in Refs.~\cite{CJ99,CJ_PLB}.

The transition form factor $F_{VP}(q^2)$ for the radiative
decay of vector meson $V(P)\to P(P')\gamma^*(q)$ is defined as \be\label{ff}
\la P(P')|J^\mu_{\rm em}|V(P,h)\ra
=ie\epsilon^{\mu\nu\rho\sigma}\epsilon_\nu(P,h)
q_{\rho}P_{\sigma}F_{VP}(q^2),
\ee
where $q=P-P'$ is the four momentum of the virtual photon and $\ep_\nu(P,h)$ is the polarization
vector~\cite{CJ07} of the vector meson with four momentum $P$ and helicity $h$.
The coupling constant $g_{VP\gamma}$ for real photon($\gamma$) case
is determined in the limit as $q^2\to 0$, i.e. $g_{VP\gamma}=F_{VP}(q^2=0)$.

We obtain the momentum dependent transition form factor $F_{VP}(q^2)$ using the Drell-Yan-West frame(
$q^+=q^0+q^3=0$)~\cite{DYW,LB} where $q^2=q^+q^- - {\bf q}^2_\perp=-Q^2$, i.e. $Q^2>0$ is the
space-like momentum transfer. In this frame, the matrix element of the current can be expressed
as convolution integral in terms of the light-front wave function without encountering zero-mode
contributions~\cite{ZM} as far as the ``$+$" component of currents $J^\mu_{\rm em}$ is used. To obtain the
time-like form factor, we analytically continue the space-like form factor $F_{VP}(Q^2)$
to the time-like($q^2>0$) region by changing $Q^2$ to $-q^2$ in the form factor.
Furthermore, we use the transverse($h=\pm 1$) polarization
to extract the coupling constant $g_{VP\gamma}$ since the longitudinal
state of the vector meson cannot convert into a real photon.

The hadronic matrix element of the plus current,
${\cal M}^+\equiv \la P(P')|J^+_{\rm em}|V(P,h=+)\ra$ in
Eq.~(\ref{ff}) is then obtained by the convolution formula of the initial
and final state light-front wave functions:
\bea\label{Jplus}
{\cal M}^+&=&\sum_{j}ee_j\int^1_0\frac{dx}{16\pi^3}
\int d^2{\bf k}_\perp\phi_R(x,{\bf k'}_\perp)
\phi_R(x,{\bf k}_\perp)
\nonumber\\
&&\times\sum_{\lam\bar{\lam}}{\cal R}^{00\dagger}_{\lam'\bar{\lam}}
\frac{\bar{u}_{\lam'}(p'_1)}{\sqrt{p'^+_1}}\gamma^+
\frac{u_{\lam}(p_1)}{\sqrt{p^+_1}}
{\cal R}^{11}_{\lam\bar{\lam}},
\eea
where ${\bf k'}_\perp={\bf k}_\perp - x_2{\bf q}_\perp$, $p^+_1=p'^+_1=x_1P^+$,
and $ee_j$ is the electrical charge for $j$-th quark flavor.
Comparing with the right-hand-side of Eq.~(\ref{ff}),
we could extract the one-loop integral $I(m_1, m_2,q^2)$ as follows~\cite{CJ07}
\bea\label{soft_form}
I(m_1, m_2,q^2) &=&\int^1_0 \frac{dx}{8\pi^3}\int d^2{\bf k}_\perp
\frac{\phi(x, {\bf k'}_\perp)\phi(x,{\bf k}_\perp)}
{x_1\tilde{M_0}\tilde{M'_0}}
\nonumber\\
&&\times
\biggl\{{\cal A}
+ \frac{2}
{{\cal M}_0}
\biggl[{\bf k}^2_\perp
-
\frac{({\bf k}_\perp\cdot{\bf q}_\perp)^2}{{\bf q}^2_\perp}\biggr]
\biggr\},
\nonumber\\
\eea
where $\tilde{M_0}=\sqrt{M^2_0-(m_1-m_2)^2}$ and ${\cal M}_0=M_0+m_1+m_2$.
The primed factors are the functions of final state momenta,
e.g. $\tilde{M}'_0=\tilde{M}'_0(x,{\bf k'}_\perp)$.

The transition form factor $F_{VP}(q^2)$ is then obtained as
\be\label{FS}
F_{VP}(q^2)= e_1I(m_1,m_2,q^2) + e_2 I(m_2,m_1,q^2),
\ee
and the decay width for $V\to P\gamma$ is given by
\be\label{width}
\Gamma(V\to P\gamma)=\frac{\alpha}{3}g_{VP\gamma}^2 k^3_\gamma,
\ee
where $\alpha$ is the fine-structure constant  and
$k_\gamma=(M^2_V-M^2_P)/2M_V$ is the kinematically allowed energy
of the outgoing photon.

In our numerical calculations, we use two sets of model parameters
($m_q=0.22, m_s=0.45,\beta_{q\bar{q}}=0.3659, \beta_{q\bar{s}}=0.3886$)[GeV] for the linear
and ($m_q=0.25, m_s=0.48,\beta_{q\bar{q}}=0.3194, \beta_{q\bar{s}}=0.3419$)[GeV] for HO confining
potentials obtained from our variational principle~\cite{CJ99,CJ_PLB}, where $q=u$ or $d$-quark.
The isospin symmetry(i.e. $m_u=m_d$) used in our LFQM
implies the relation of the transition form factors $F_{\omega\pi}(Q^2) =3F_{\rho\pi}(Q^2)$ between
$\rho\to\pi\gamma^*$ and $\omega\to\pi\gamma^*$ processes.
In Ref.~\cite{CJ99}, we have shown that weak decay constants and electromagnetic charge radii of
($\pi,K,\rho,\omega, K^*$) mesons as well as radiative decay widths for $\rho(\omega)\to\pi\gamma$
and $K^*\to K\gamma$ are quite comparable with the experimental data.

\begin{figure}
\vspace{0.9cm}
\includegraphics[width=2.8in,height=2.8in]{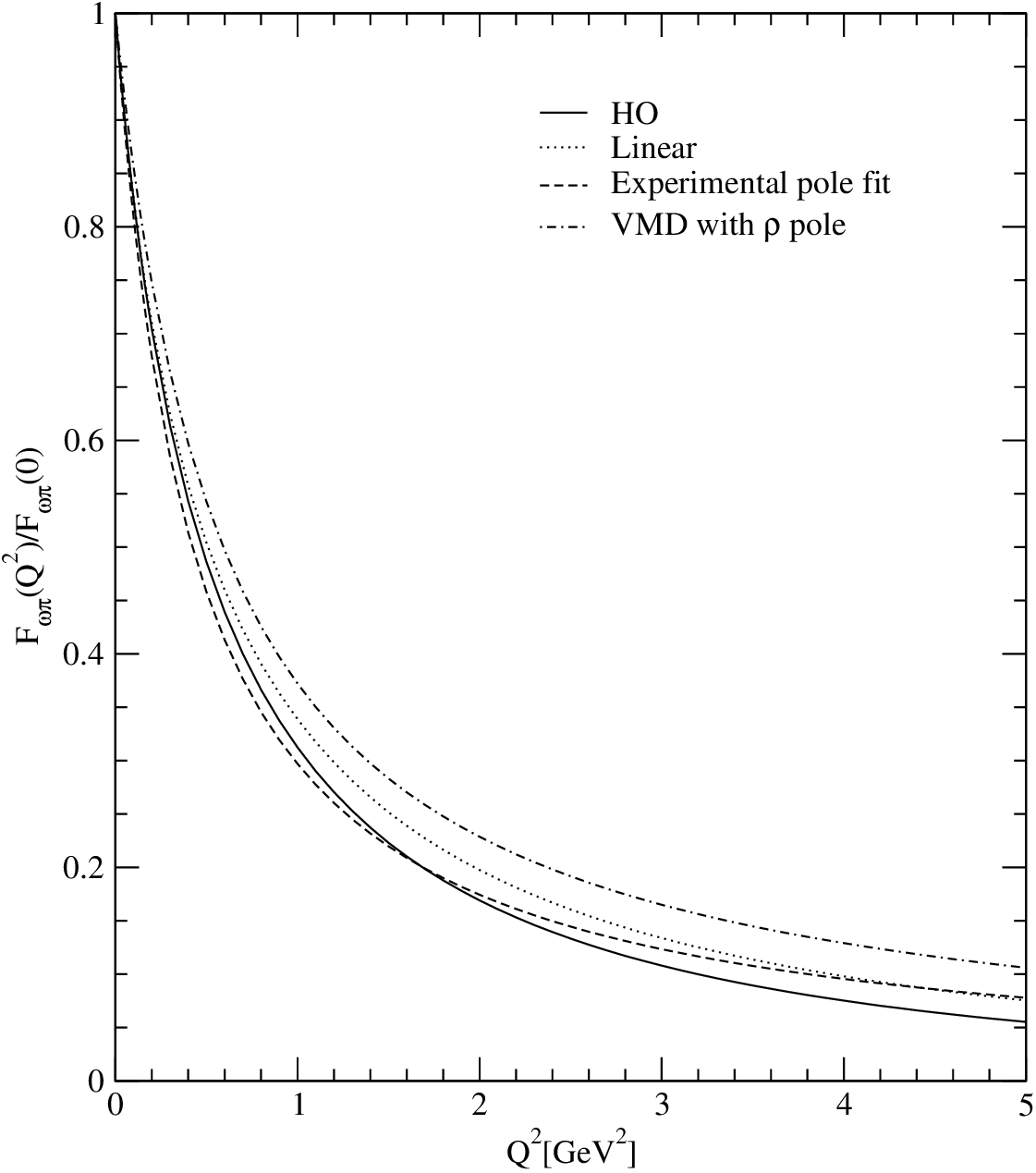}
\caption{The normalized transition form factor of
$\omega\to\pi^0\gamma^*$(or $\rho^{\pm}\to\pi^{\pm}\gamma^*$) in space-like $Q^2$ region obtained
from HO(solid line) and linear(dotted line) potential models compared with experimental pole fit(dashed line)
as well as VMD model(dot-dashed line).}
\label{fig1}
\end{figure}
In Fig.~\ref{fig1}, we show our results of the normalized transition form factor
$F_{\omega\pi}(Q^2)/F_{\omega\pi}(0)$[or $F_{\rho^{\pm}\pi^{\pm}}(Q^2)/F_{\rho^{\pm}\pi^{\pm}}(0)$]
for the $\omega\to\pi\gamma^*$[or $\rho^{\pm}\to\pi^{\pm}\gamma^*$] transition as a function
of the photon momentum $Q^2$. The solid and dotted lines represent the results of our HO and linear
potential models, respectively. The dashed and dot-dashed lines represent the results of experimental
pole fit, $F^{pole}_{\omega\pi}(Q^2)=1/(1+Q^2/(\Lambda^{\rm exp}_{\omega\pi})^2)$ with
$\Lambda^{\rm exp}_{\omega\pi}=0.65$ GeV and vector meson dominance(VMD) model with $\rho$ pole,  $F^{VMD}_{\omega\pi}(Q^2)=1/(1+Q^2/(\Lambda^{VMD}_{\omega\pi})^2)$ with
$\Lambda^{VMD}_{\omega\pi}=0.77$ GeV, respectively. We have shown in~\cite{CJ99} that
our results for the coupling constants $g_{\omega\pi\gamma}=2.349[2.242]$ GeV$^{-1}$
and $g_{\rho\pi\gamma}=0.783[0.747]$ GeV$^{-1}$ obtained from HO[linear] model were in good agreement
with the experimental data $g^{\rm Exp.}_{\omega\pi\gamma}=(2.319\pm 0.083)$ GeV$^{-1}$ and
$g^{\rm Exp.}_{\rho\pi\gamma}=(0.733\pm 0.038)$ GeV$^{-1}$~\cite{PDG}. As one
can see from Fig.~\ref{fig1}, the momentum dependent form factors obtained from both HO and linear models
are also quite close to  $F^{pole}_{\omega\pi}(Q^2)$ at least for small $Q^2$ region.
We also obtain the electromagnetic radius of the form factor
$F_{\omega\pi}(Q^2)$ as $\la r^2_{\omega\pi}\ra= 1.199[1.183]$ fm$^2$ for HO[linear] potential
model, which can be compared  with 0.897 fm$^2$ from other quark model calculation~\cite{YXM}.

\begin{figure}
\vspace{0.9cm}
\includegraphics[width=2.8in,height=2.8in]{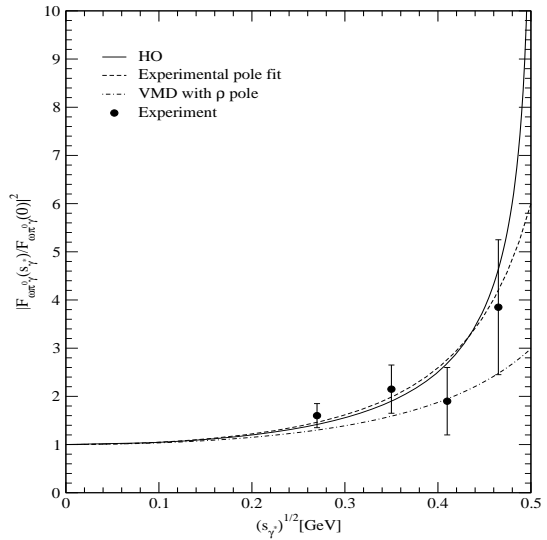}
\caption{Time-like $\omega\to\pi\gamma^*$ transition form factor obtained from HO potential model(solid line)compared with experimental pole fit(dashed line) and VMD model(dot-dashed line).}
\label{fig2}
\end{figure}
In Fig.~\ref{fig2}, we show the time-like form factor of
$\omega\to\pi\gamma^*$ obtained from our HO model(solid line) and compare with the experimental
data~\cite{Dz} as well as $F^{pole}_{\omega\pi}(q^2)$ (dotted line) and $F^{VMD}_{\omega\pi}(q^2)$
(dot-dashed line). Our result for the time-like form factor is in good agreement
with the data. We should note that we only give the results below the particle production
threshold($q^2=4m^2_q$) since the singularity for bound state production and the imaginary part
will appear beyond the threshold in our model calculation.

\begin{figure}
\vspace{0.9cm}
\includegraphics[width=2.8in,height=2.8in]{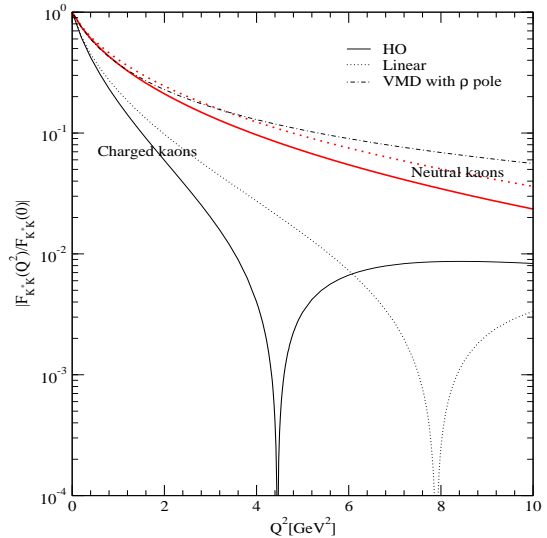}
\caption{(color online). The normalized neutral(red) and
charged(black) $K^*K\gamma^*$ form factors in space-like region obtained from HO(solid line) and
linear(dotted line) potential models compared with VMD model(dot-dashed line).}
\label{fig3}
\end{figure}

The transition form factors for charged $K^{*\pm}\to K^{\pm}\gamma^*$ and neutral $K^{*0}\to K^0\gamma^*$
processes are quite interesting quantities as the couplings of the two currents to the quark and the antiquark differ
because of the $SU(3)$ flavor symmetry breaking. In Fig.~\ref{fig3}, we show the normalized neutral(red online) and
charged(black online) $K^*K\gamma^*$ form factors in space-like region obtained from our HO(solid line) and linear(dotted line) potential models, respectively. Our LFQM results are also compared with
VMD model(dot-dashed line). While the momentum dependent behaviors of $F_{K^{*0}K^{0}}(Q^2)$ show a nearly VMD-like behavior at least for small $Q^2$ region, those of $F_{K^{*\pm}K^{\pm}}(Q^2)$
are very different from the VMD result. Especially, our $F_{K^{*\pm}K^{\pm}}(Q^2)$ encounters zero
at $Q^2\simeq 4.5$ GeV$^2$ for HO model and $Q^2\simeq 8$ GeV$^2$ for linear model, respectively.
The authors in~\cite{Munz} also found form factor zero using the covariant Bethe-Salpeter(BS) model, where
the zero occurs at $Q^2= 4.8$ GeV$^2$ for $m_s/m_q=1.8$ and moves to the right(left) as $m_s/m_q$
decreases(increases). In our model calculation, the point of form factor zero moves to the right as $m_s/m_q$
decreases for given $\beta$ or as $\beta$ increases for given $m_s/m_q$ and vice versa.
The form factor zero for $K^{*\pm}\to K^{\pm}\gamma^*$ decay is mainly due to the negative interference
between the two currents and depends sensitively on the ratio of the mass of the strange and nonstrange constituent quarks, i.e. $m_s/m_q$ as well as the gaussian $\beta$ parameters.

\begin{figure}
\vspace{0.9cm}
\includegraphics[width=2.8in,height=2.8in]{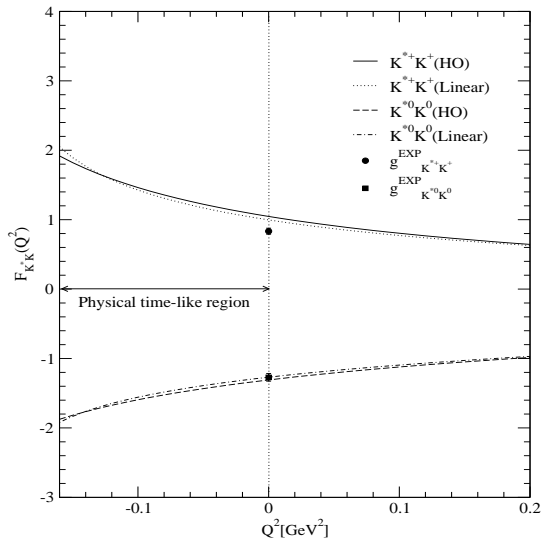}
\caption{The space- and time-like $K^{*+}K^{+}\gamma^*$ and $K^{*0}K^0\gamma^*$ form factors obtained from HO
and linear potential models for $-0.16\leq Q^2\leq 0.2$ GeV$^2$ region.}
\label{fig4}
\end{figure}
In Fig.~\ref{fig4}, we show both space- and time-like form factors of $K^{*+}\to K^{+}\gamma^*$ and $K^{*0}\to K^0\gamma^*$ transitions obtained from HO and linear potentials for $-0.16\leq Q^2\leq 0.2$ GeV$^2$ region, where
$q^2_{\rm max}=(M_{K^*}-M_K)^2\simeq 0.16$ GeV$^2$ and $q^2=0$ correspond to a final state $K$ meson recoiling with zero and maximum three-momentum, respectively. The line codes are explained in the figure.
While our value of the coupling constant
$g_{K^{*+}K^+\gamma}=1.047[0.997]$ GeV$^{-1}$ obtained from HO[linear] model is slightly larger than the experimental data
$g^{\rm exp}_{K^{*+}K^+\gamma}=(0.834\pm 0.041)$ GeV$^{-1}$~\cite{PDG}, $g_{K^{*0}K^0\gamma}=-1.309[-1.269]$ GeV$^{-1}$ obtained
from HO[linear] is in good agreement with the data $g^{\rm Exp.}_{K^{*0}K^0\gamma}=-(1.271\pm 0.055)$ GeV$^{-1}$~\cite{PDG}.
The deviation of the coupling constant ratio of $|g_{K^{*0}K^0\gamma}/g_{K^{*+}K^+\gamma}|$ from 2 implies the amount of $SU(3)$ symmetry breaking effect~\cite{BS,Jones}. Although our HO and linear model predictions for both neutral and charged kaon decays are somewhat different from each other in the intermediate and deep space-like $Q^2$ region(see Fig.~\ref{fig3}), the two models are not much differ for the physical time-like region as well as the small $Q^2$ region.

In this Brief Report, we investigated the magnetic dipole $\omega\to\pi\gamma^*$ and $K^*\to K\gamma^*$ transitions using our LFQM constrained by the variational principle for the QCD-motivated effective
Hamiltonian.
The momentum dependent form factors $F_{VP}(q^2)$ for $V\to P\gamma^*$ decays
are obtained in the $q^+=0$ frame and then analytically continued
to the time-like region by changing ${\bf q}_\perp$ to $i{\bf q}_\perp$ in the form factors.
The coupling constants $g_{VP\gamma}$ for real photon case is determined in the limit
as $q^2\to 0$, i.e.  $g_{VP\gamma}=F_{VP}(q^2=0)$. One of the features we have investigated is the finding of
the form factor zero for charged $K^{*\pm}\to K^{\pm}\gamma^*$ transition, which deserves special attention from the viewpoint of experimental possibility.
As a concluding remark, our model parameters obtained from the variational principle uniquely determine the above nonperturbative quantities. This approach can establish the extent of applicability of our
LFQM to wider ranging hadronic phenomena.

This work was supported by Kyungpook National University Research Fund, 2007.

\end{document}